\documentstyle[11pt]{article}
\renewcommand{\baselinestretch}{1.75} 
\def\text{}
\def\cF{{\cal F}}
\def\cl{{\rm cl}}
\def\NSzero{ {\bigcirc \!\!\!\!\! 0}\,\,}
\def\NSone{ {\bigcirc \!\!\!\!\! 1}\,\,}
\def\NStwo{ {\bigcirc \!\!\!\!\! 2}\,\,}
\def\NSthree{ {\bigcirc \!\!\!\!\! 3}\,\,}
\def\NSfour{ {\bigcirc \!\!\!\!\! 4}\,\,}
\def\NSfive{ {\bigcirc \!\!\!\!\! 5}\,\,}
\def\e{{\rm e}}
\def\oneloop{{\rm 1-loop}}
\def\oneinst{{\rm 1-inst}}

\def\threeinst{{\rm 3-inst}}
\def\dinst{{d-\rm{inst}}}
\def\half{ {\textstyle{1\over 2}}}
\def\quarter{ {\textstyle{1\over 4}}}

\def\Aone{ A^{(1)} }
\def\Atwo{ A^{(2)} }
\def\Athree{ A^{(3)} }
\def\Afour{ A^{(4)} }
\def\Afive{ A^{(5)} }
\def\Asix{ A^{(6)} }

\def\Bone{ B^{(1)} }
\def\Btwo{ B^{(2)} }
\def\Bthree{B^{(3)} }
\def\Bfour{ B^{(4)} }
\def\Bfive{ B^{(5)} }
\def\Bsix{  B^{(6)} }
\newcommand{\La}{\Lambda}
\newcommand{\th}{\theta}
\newcommand{\la}{\lambda}
\newcommand{\be}{\begin{eqnarray}}
\newcommand{\ee}{\end{eqnarray}}
\newcommand{\pr}{\partial} 
\newcommand{\np}{\newpage}
\newcommand{\hs}{\hspace}
\newcommand{\vs}{\vspace}

\newcommand{\nn}{\nonumber}
\newcommand{\lra}{\longrightarrow}
\newcommand{\llra}{\longleftrightarrow}

\begin{document}
\thispagestyle{empty}
\vs*{-25mm}
\begin{flushright}
BRX-TH-447\\[-.2in]
BOW-PH-114\\[-.2in]
HUTP-98/A085 \\[-.2in]
hep-th/9904078
\end{flushright}

\begin{center}
{\LARGE{\bf 
Two antisymmetric hypermultiplets in\\[-.2in]
${\cal N}$=2 SU($N$) gauge theory:\\
Seiberg-Witten curve and M-theory interpretation
}} \\
\vspace{.2in}

\renewcommand{\baselinestretch}{1}
\small
\normalsize
Isabel P. Ennes\footnote{Research supported 
by the DOE under grant DE--FG02--92ER40706.}\\
Martin Fisher School of Physics\\
Brandeis University, Waltham, MA 02454

\vspace{.1in}
Stephen G. Naculich\footnote{Research supported 
in part by the National Science Foundation under grant no. PHY94-07194.}\\
Department of Physics\\
Bowdoin College, Brunswick, ME 04011

\vspace{.1in}

Henric Rhedin\\
Department of Engineering Sciences,
 Physics and Mathematics\\
Karlstad University, S-651 88 Karlstad, Sweden\\
\vspace{.1in}

Howard J. Schnitzer\footnote{Research supported in part
by the DOE under grant DE--FG02--92ER40706.}\\
Martin Fisher School of Physics\footnote{Permanent address.\\
{\tt \phantom{aaa} naculich@bowdoin.edu; henric.rhedin@kau.se;
ennes,schnitzer@binah.cc.brandeis.edu}}\\
Brandeis University, Waltham, MA 02454\\
and\\
Lyman Laboratory of Physics\\
Harvard University, Cambridge, MA 02138\\

\vspace{.2in}

{\bf{Abstract}} 
\end{center}
\renewcommand{\baselinestretch}{1.75}
\small
\normalsize
\begin{quotation}
\baselineskip14pt
\noindent 

The one-instanton contribution to the prepotential 
for ${\cal N}=2$ supersymmetric gauge theories
with classical groups 
exhibits a universality of form. 
We extrapolate the observed regularity 
to SU$(N)$ gauge theory with two antisymmetric hypermultiplets
and $N_f\leq 3$ hypermultiplets in the defining representation. 
Using methods developed for the instanton expansion 
of non-hyperelliptic curves, 
we construct an effective quartic Seiberg-Witten curve 
that generates this one-instanton prepotential. 
We then interpret this curve in terms of an M-theoretic picture
involving NS 5-branes, D4-branes, D6-branes, and orientifold sixplanes,  
and show that for consistency, an 
infinite chain of 5-branes and orientifold sixplanes is required,
corresponding to a curve of infinite order.
\end{quotation}

\np 

\setcounter{page}{1}
\setcounter{footnote}{0}
\noindent{\bf 1. Introduction}
\renewcommand{\theequation}{1.\arabic{equation}}
\setcounter{equation}{0}

The program of Seiberg and Witten \cite{SeibergWitten}
allows one to extract the exact low-energy behavior 
of four-dimensional ${\cal N}=2$ supersymmetric gauge 
theories from the following data: 
a Riemann surface or algebraic curve 
specific to the group and representation content 
of the underlying Lagrangian, 
and a preferred meromorphic 1-form, the Seiberg-Witten (SW) differential. 
{}From this information, one may (in principle) reconstruct 
the prepotential of the Coulomb branch of the theory 
in the low-energy limit 
from the period integrals of the SW differential. 
In practice, technical difficulties make the construction
of the prepotential a challenging problem.

The Riemann surface associated with
${\cal N}=2$ supersymmetric gauge theories based on the classical groups,
either without matter hypermultiplets
or with hypermultiplets in the defining representation, 
is hyperelliptic \cite{Everybody}. 
In this case systematic methods are available for extracting 
the relevant physical information 
\cite{DHokerKricheverPhong1}--\cite{IsidroEdelsteinMarinoMas}.

For other ${\cal N}=2$ supersymmetric gauge theories, however,
the associated Riemann surface, 
obtained via geometric engineering \cite{geomengineering,productgroupcurve} 
or M-theory \cite{Mtheory}--\cite{LandsteinerLopezLowe},
is not hyperelliptic, 
and in fact one may encounter varieties 
that are not Riemann surfaces at all. 
For SU($N$) gauge theories 
with one hypermultiplet in the symmetric (or antisymmetric) representation
(with or without additional hypermultiplets in the defining representation),
the Riemann surface is described 
by a cubic (non-hyperelliptic) curve 
\cite{LandsteinerLopezLowe}. 
For a gauge theory based on a product of $m$ factors of SU($N$)
with hypermultiplets in bifundamental representations,
the Riemann surface is described by 
an $(m+1)$th order curve \cite{Witten,productgroupcurve}.
In a series of papers 
\cite{oneanti}--\cite{product}, 
we have developed a systematic approximation scheme 
to compute the instanton expansion of the prepotential 
for non-hyperelliptic curves of cubic and higher order.
This allows one to test the predictions of M-theory and geometric engineering
for field theory,
thereby increasing our confidence in the validity of these
string-theoretic methods.

In this paper we will discuss the SW problem for SU($N$)  gauge theory
with two matter hypermultiplets in the antisymmetric representation 
and up to 3 additional hypermultiplets in the defining representation. 
Since there is no existing M-theoretic
or geometric engineering prediction for the curve for this theory,
our methods will differ from previous work on this subject. 
We begin by {\it predicting} the form of $\cF_{\oneinst}$ 
from the observed regularities of known prepotentials in section 2.

In section 3,
we then ``reverse engineer'' a Seiberg-Witten curve for this theory
from the prepotential,
using methods we have developed \cite{oneanti}--\cite{product} 
for computing the instanton expansion for non-hyperelliptic curves.
The quartic curve that we derive has the correct limiting behavior 
as the mass of either of the antisymmetric hypermultiplets goes to infinity.

We then attempt an M-theory interpretation\footnote{We 
describe the brane structure in terms of type IIA string theory, 
which then lifts to M-theory~\cite{Witten}.}
of the result in section 4. 
The quartic curve corresponds to a picture 
containing four parallel NS 5-branes, 
with each adjacent pair linked by $N$ D4-branes,
and four orientifold 6-planes, one on each 5-brane. 
Hypermultiplets in the defining representation 
correspond to additional D6-branes. 
The reflection symmetries of the orientifold 6-planes,
however, imply an expanded M-theory picture
with an infinite chain of equally spaced  parallel NS 5-branes, 
and an infinite set of orientifold 6-planes,
one lying on each of the NS 5-branes. 
Thus, the effective quartic curve derived in section 3 is only a
truncation of a curve of infinite order.
To calculate the prepotential to any given order in the instanton
expansion,  however, the curve corresponding to  only a finite subset 
of 5-branes is needed.
For example, $2d$ NS 5-branes
(corresponding to a curve of order $2d$) are necessary
to compute the prepotential to $\half d(d-1)$-instanton accuracy.

In section 5 we sum the infinite series representing 
the leading order coefficients in the curve for certain special cases, 
and are able to represent the curve in terms of theta functions. 
This leads us to speculate that our curve 
is related to a ``decompactification" of the elliptic model described 
in M-theory by Uranga \cite{Uranga}, 
who considered the scale invariant case of SU$(N)$ 
with two hypermultiplets in the antisymmetric representation 
and four hypermultiplets in the defining representation
(although he does not specify a curve for this theory).
The link to our work would be if one (or more) 
of the defining hypermultiplets had their mass(es) sent to infinity.

Section 6 summarizes our results, and points to issues for further study.

\bigskip
\noindent{\bf 2. The Prepotential}
\renewcommand{\theequation}{2.\arabic{equation}}
\setcounter{equation}{0}

The Lagrangian for an ${\cal N}=2$ gauge theory 
to lowest order in the momentum expansion is  
\be
{\cal L}=\frac{1}{4\pi}{\rm Im}\left[\int{\rm d}^4\th
\frac{\pr\cF(A)}{\pr A^i}\bar{A}^i+
\frac{1}{2}\int{\rm d}^2\th \frac{\pr^2\cF(A)}{\pr A^i\pr A^j}
W^iW^j\right],
\label{one}
\ee
where $A^i$ are ${\cal N}=1$ chiral superfields. 
The prepotential in the Coulomb phase has the form 
\be
\cF(A) = \cF_{\cl}(A)
        +\cF_{\oneloop}(A)
	+ \sum_{d=1}^{\infty}\La^{[I(G)-I(R)]d}\cF_{\dinst}(A),
\label{two}
\ee
where $I(G)$ [$I(R)$] is the Dynkin index of the 
adjoint (matter) representation. 
The prepotential (\ref{two})
may be obtained from the Seiberg-Witten data by first computing
the renormalized order parameters and their duals 
\be
2\pi i a_k=\oint_{A_k}\lambda 
\hs{10mm} {\rm and} \hs{10mm}
2\pi i a_{D,k}=\oint_{B_k}\lambda,
\label{three}
\ee
where $\lambda$ is the Seiberg-Witten differential and
$A_k$ and $B_k$ are a canonical basis of homology cycles
for the Riemann surface, 
and then integrating 
\be
a_{D,k}=\frac{\pr\cF}{\pr a_k}. 
\label{four}
\ee

The one-instanton contribution to the prepotential, $\cF_{\oneinst}$,
for the classical groups exhibits a remarkable universality of form when 
expressed in terms of the renormalized order parameters. 
In particular, 
for SU$(N)$, the one-instanton prepotential has the form 
\cite{DHokerKricheverPhong1,onesym,nonhyper}
\be
8\pi i\,\cF_{\oneinst}=\, \sum_{k=1}^{N}\, S_k\,(a_k), 
\label {five}
\ee
for 
(a) $N_f$ hypermultiplets in the defining representation, or 
(b) one hypermultiplet in the symmetric representation 
and $N_f$ hypermultiplets in the defining representation, 
while for
(c) one hypermultiplet in the antisymmetric representation
and $N_f$ hypermultiplets in the defining representation,  
it is given by 
\cite{oneanti,nonhyper}
\be
8\pi i\,\cF_{\oneinst}=\, \sum_{k=1}^{N}\, S_k\,(a_k)\,-\,2\,S_m\,(-m). 
\label {six}
\ee
The expressions $S_k (a_k)$ and $S_m (-m)$ 
in eqs.~(\ref{five}) and (\ref{six})
are the coefficients of second order poles 
at $x=a_k$ and $x=-m$, respectively,
of a function $S(x)$,
\be
S(x) =  {S_k (x) \over (x-a_k)^2 } =   {S_m (x) \over (x + m)^2 },
\label{sixa}
\ee
which can be obtained from the hyperelliptic approximation 
to the curve, if it is known. 
The explicit form of $S(x)$ for each of the above theories
is specified in Table 1.
(The relative coefficient between the terms in (\ref{six}) 
guarantees the absence of a pole in $\cF_{\oneinst}$ at $a_k = -m$.)

One observes from the table the following regularities 
for $S(x)$, 
which is a product of factors:

\noindent
(1) a factor 
\be
{1\over {\prod_{i=1}^{N}\,(x-a_i)^2}}\,,
\label {seven}
\ee
for the pure gauge multiplet, 

\noindent
(2) a factor 
\be
(x+M_j)\,, 
\label {eight} 
\ee
for each hypermultiplet of mass $M_j$ in the defining representation, 

\noindent
(3) a factor 
\be
(-1)^N\,(x+m)^2\,\prod_{i=1}^{N}\,(x+a_i+2m)\,, 
\label {nine}
\ee
for a hypermultiplet of mass $2m$ in the symmetric representation, and

\begin{center}
\begin{tabular}{||c|c||}
\hline\hline
Hypermultiplet Representations &  $S(x)$\\
\hline\hline 
	&{}                            					\\ 
$N_f$ defining                     
	& ${4\prod_{j=1}^{N_f}(x+M_j)\over \prod_{i=1}^N (x-a_i)^2}$	\\
(ref. \cite{DHokerKricheverPhong1})       	 
	&{}								\\
\hline
1\,\,{\rm symmetric}  
        &{}								\\
$+ N_f \,\,{\rm defining}$ 
	& $ {4(-1)^N(x+m)^2
	     \prod_{i=1}^N (x+a_i+2m) \prod_{j=1}^{N_f}(x+M_j) 
     		\over \prod_{i=1}^N (x-a_i)^2}$ 	\\
(ref. \cite{onesym,nonhyper})		
    	&{}								\\
\hline
1\,\,{\rm antisymmetric}	
	&{}								\\ 
$+N_f\,\,{\rm defining}$   
	& ${4(-1)^N  
	\prod_{i=1}^N (x+a_i+2m) \prod_{j=1}^{N_f}(x+M_j) 
	\over (x+m)^2 \prod_{i=1}^N (x-a_i)^2}$			\\
(ref. \cite{oneanti,nonhyper})
	&   								 \\
\hline
2 \,\,{\rm antisymmetric}               
	&{}								\\
$ +N_f \,\,{\rm defining}$	       
	& ${4\prod_{i=1}^{N}(x+a_i+2m_1) \prod_{i=1}^N(x +a_i+2m_2)
		\prod_{j=1}^{N_f}( x +M_j) \over ( x +m_1)^2 ( x +m_2)^2
                 \prod_{i=1}^N ( x -a_i)^2}$				\\
(This paper.)				 
	&{}								\\
\hline

\end{tabular}
\end{center}
\label{tableone}
{\footnotesize{\bf Table 1}: The function $S(x)$
for SU$(N)$ gauge theory, with various matter contents.
The hypermultiplets in the defining representation have masses $M_j$. 
The symmetric or antisymmetric representation has mass $2m$. 
If there are two antisymmetric representations, 
their masses are $2m_1$ and $2m_2$}.
\vspace{1.5cm}

\noindent
(4) a factor 
\be
{(-1)^N\over (x+m)^2}\,
\prod_{i=1}^{N}\,(x+a_i+2m)\,, 
\label {ten}
\ee
for a hypermultiplet of mass $2m$ in the antisymmetric representation.

The first three entries of Table 1 almost exhaust the (generic) 
cases for the Coulomb phase of 
${\cal N}=2$ supersymmetric SU$(N)$ gauge theories. 
The remaining (generic) case 
of two antisymmetric hypermultiplets
(with up to 3 additional hypermultiplets in the defining representation) 
has not been treated to date. 
Since we know of no M-theoretic or geometric engineering description 
of this case, 
we begin instead by {\it predicting} $\cF_{\oneinst}$,
and then reverse engineer the SW curve.
Based on the regularities described above,
we postulate in the last row of Table 1 the form of $S(x)$ for 
two antisymmetric hypermultiplets with masses $2m_1$ and $2m_2$ and 
$0 \le N_f \le 3$  hypermultiplets in the defining 
representation with masses $M_j$.
We then predict the one-instanton prepotential to have the form
\be
8\pi i\,\cF_{\oneinst}=\, \sum_{k=1}^{N}\, 
S_k\,(a_k)\,-\,2\,S_{m_1}\,(-m_1) \,-\, 2\,S_{m_2}\,(-m_2) \,+\, C,
\label{eleven}
\ee
where, as before,
\be
S(x) =  {S_k (x) \over (x-a_k)^2 } 
     = {S_{m_1} (x) \over (x + m_1)^2 }
     = {S_{m_2} (x) \over (x + m_2)^2 }.
\label{elevena}
\ee
Explicitly,
\be
S_k (a_k) 
& = & 
{4\prod_{i=1}^{N}(a_k+a_i+2m_1)(a_k +a_i+2m_2)
\prod_{j=1}^{N_f}(a_k +M_j) \over (a_k+m_1)^2 ( a_k +m_2)^2
\prod_{i\ne k}^N (a_k -a_i)^2}	,
\nn \\ [.2in]
S_{m_1} (-m_1)
& = &
{4 \prod_{i=1}^N(a_i+2m_2-m_1)
\prod_{j=1}^{N_f}(M_j-m_1) \over (m_2-m_1)^2
\prod_{i=1}^N (a_i+m_1)},
\label{elevenaa} \\ [.2in]
S_{m_2} (-m_2)
& = & 
{4\prod_{i=1}^{N}(a_i+2m_1-m_2)
\prod_{j=1}^{N_f}( M_j-m_2) \over (m_1-m_2)^2 
\prod_{i=1}^N (a_i+m_2)}.
\nn 
\ee
As before, the relative coefficients in the prepotential (\ref{eleven})
guarantee the absence of poles at $a_k = -{m_1}$ and $a_k = -{m_2}$.
The inclusion of the constant
\be
C = {16\over (m_2-m_1)^2}  \prod_{j=1}^{N_f} \left (M_j- \half[m_1+m_2]\right)
\ee
in eq.~(\ref{eleven}), 
although irrelevant to the computation of the dual order parameters,
renders the prepotential finite in the limit $m_2 \to m_1$.

One can test the postulated form of the prepotential (\ref{eleven}) 
by considering the special cases:

\noindent
(a) $N=2$, which is equivalent to SU(2) gauge theory
with $N_f\leq 3$ defining hypermultiplets.

\noindent
(b) $N=3$, which is equivalent to 
SU(3) gauge theory with 2 anti-defining 
and $N_f\leq 3$ defining hypermultiplets, 
or equivalently, SU(3) with $2\leq N_f\leq 5$ defining hypermultiplets. 

\noindent
(c) The limit $m_1$ or $m_2 \to \infty$, 
which removes one of the antisymmetric hypermultiplets from the theory,
in which case eq. (\ref{eleven}) should reduce to eq.~(\ref{six}).

In each of these cases, there is complete agreement
(up to an irrelevant constant). 
Therefore, for the remainder of the paper we take 
eqs. (\ref{eleven}) and (\ref{elevenaa})
to correctly describe the one-instanton prepotential 
for SU($N$) gauge theory with
two antisymmetric and $N_f\leq 3$ defining hypermultiplets.

Finally, we differentiate the prepotential to obtain
the dual order parameter $a_{D,k}$ for this theory.
Using eq.~(\ref{eleven}) together with the one-loop 
contribution to the prepotential, given by perturbation theory,
\be
\cF_{\oneloop} 
 =   \frac{i}{8\pi}
&\bigg[&
\sum^N_{i,j=1} (a_i-a_j )^2 \log \frac{(a_i-a_j)^2}{\La^2 }
\; - \; 
\sum_{j=1}^{N_f}
\sum_{i=1}^N 
(a_i + M_j)^2 \log \frac{(a_i + M_j)^2}{\La^2 } 
\nn\\ [.1in]
&& 
- \; \sum_{\ell=1}^2 
\sum_{i < j}^N 
(a_i + a_j + 2m_\ell)^2 \log \frac{(a_i + a_j+2m_\ell)^2}{\La^2 } 
\bigg]\!,\,\,\,\,\,
\label{elevenb}
\ee
we use (\ref{two}) and (\ref{four}) to find
\be
2 \pi i a_{D,k}
& = &
[{\rm const}] a_k  
-  2  \sum_{i\neq k}^N (a_k-a_i) \log (a_k-a_i)  
+     \sum_{j=1}^{N_f} (a_k+M_j)  \log (a_k+M_j) 
\nn\\ [.1 in]
&+&    \sum_{\ell=1}^2   
     \left[ \sum_i (a_k+a_i+2m_\ell)  \log (a_k+a_i+2m_\ell) 
   -  2 (a_k+m_\ell) \log (a_k+m_\ell)  \right] \nn \\ [.2 in]
&+& \half \La^{4-N_f} \bigg[ \half \frac{\partial S_k}{\partial x} (a_k) \;
	- \sum_{i\neq k}^N \frac{S_i(a_i) }{a_k-a_i}  
	+ \half \sum_{\ell=1}^2 \sum^N_{i=1} \frac{S_i(a_i)}{a_k+a_i+2m_\ell}
			\label{elevenc}	\\ [.1 in]
&+&       \frac{S_{m_1} (-m_1)}{a_k+m_1}
   	- \frac{S_{m_1} (-m_1)}{a_k+2m_2-m_1}
   	+ \frac{S_{m_2} (-m_2)}{a_k+m_2}
   	- \frac{S_{m_2} (-m_2)}{a_k+2m_1-m_2} \bigg] + O(\La^{8-2N_f}),
\nn
\ee
accurate to one-instanton order. 
In the next section, 
we will reproduce this expression from the period
integrals of a Riemann surface.

\vfil\break
\noindent{\bf 3. The Curve}
\renewcommand{\theequation}{3.\arabic{equation}}
\setcounter{equation}{0}

Beginning from the one-instanton prepotential (\ref{eleven})
for SU($N$) gauge theory with
two antisymmetric and $N_f\leq 3$ defining hypermultiplets
postulated in the last section, 
we will now ``reverse engineer'' an effective SW curve that can
generate this prepotential.

The SW curve associated with SU($N$) 
gauge theory with matter hypermultiplets in the defining representation 
is quadratic \cite{SeibergWitten,Everybody}
while for the theory with one hypermultiplet in the symmetric or antisymmetric
representations, the curve is cubic \cite{LandsteinerLopezLowe}.
We expect at least a cubic curve for two antisymmetric hypermultiplets.
The curve for a theory with a product of $m$ factors of SU($N$) 
with matter in  bifundamental representations \cite{Witten,productgroupcurve} 
is of order $m+1$, 
but we found \cite{product} that 
to compute the dual order parameters for any of the factor groups 
to one-instanton accuracy, 
it suffices to use a quartic approximation to the full curve. 
(In all cases, the quadratic approximation 
is sufficient to compute the one-loop prepotential.)

For these reasons, we postulate a quartic curve for the theory
with two hypermultiplets in the antisymmetric representation
and $N_f\leq 3$, which takes the general form of 
the curve of ref.~\cite{Witten},
\be
   & & L^4 \, j_{1}(x) 	P_{ 2} (x)  			\, t^2 
\,+\,  L   \,           P_{ 1} (x) 			\, t
\ +\          		P_{ 0} (x)          			\   		\cr 
& &  +\ L      j_0(x)    P_{-1} (x)         		\, t^{-1}
\ +\   L^4 \, j_0^2(x)  j_{-1} (x)   P_{-2} (x)   	\, t^{-2}  
\,=\,0,
\label{twelve}
\ee 
where 
$ L^2 = \Lambda^{4-N_f}$
and the coefficient functions $P_n(x)$ and $j_n(x)$
are to be determined below.\footnote{Alternatively, 
the factors of $j_0(x)$ can be associated 
with the positive powers of $t$ through the change of variables 
$t \to t j_0(x)$,
or more controversially, distributed symmetrically between positive
and negative powers of $t$ through $t \to t j_0^{1/2}(x)$.} 
(The $j_n(x)$ are written separately from the $P_n(x)$ 
to represent the contribution of the
$N_f$ hypermultiplets in the defining representation.)
The coefficient functions are chosen to satisfy
\be
P_n (x;m_1,m_2)&=&P_{-n}(x;m_2,m_1)\,,\nn \\
j_n (x;m_1,m_2)&=&j_{-n}(x;m_2,m_1)\,,
\label{fifteen}
\ee
so that the curve is invariant under an involution that exchanges  
the two antisymmetric hypermultiplets
\be
t \lra {j_0(x) \over t}
\hs{.3in} ; \hs{.3in}
m_1\leftrightarrow m_2.
\label{fourteen}
\ee

We begin by changing variables 
\be
t\,={y\over L P_{1}(x)}, 
\label{sixteen}
\ee
to recast the curve into a form
suitable for the hyperelliptic expansion  \cite{oneanti}-\cite{product}
\be
       		{L^2 j_1(x) \,P_{2} (x)\over P_{1}^2(x)} \,     y^4
\ + \    					   		y^3
& + &      					P_0(x)     \,	y^2
								\nn\\[.1in]
		\ + \  L^2\,  j_0(x) \, P_{1}(x) \, P_{-1}(x)        \,	y 
& + &        	 L^6\, j_0^2(x)\,j_{-1}(x) \, P_{1}^2(x)\,P_{-2}(x)
\,=\, 0.
\label{seventeen}
\ee
The first approximation to eq.~(\ref{seventeen}) in an
instanton expansion is the hyperelliptic curve
\be
y^2\,+\,2 A (x) y\,+\,B(x)\,=\,0,
\label{eighteen}
\ee
where 
\be
A(x) = \half P_0(x) 
\hs{.5in} {\rm and} \hs{.5in}
B(x) = L^2 \, j_0(x) \, P_{1}(x)\,P_{-1}(x).
\label{tone}
\ee
We develop a systematic expansion about the hyperelliptic approximation
(\ref{eighteen}), 
where on one of the sheets of the Riemann surface (\ref{seventeen}),
\be
y \,=\, y_I \,+\, y_{II} \,+\ \cdots  \,,
\label{tsix}
\ee
with
\be
y_I = -A-r   \hs{.5in} {\rm and} \hs{.5in}    r=\sqrt{A^2-B},
\label{tsixa}
\ee
the solution to the hyperelliptic approximation (\ref{eighteen}), 
and \cite{product}
\be
y_{II} \  =\  -\ {(A+r)^3\over 2r} 
       {L^2\,j_{ 1}(x) \,P_{2}(x)  \over P_{1}^2(x) } 
\ - \ {1 \over 2r (A+r)} L^6\, j_0^2(x) \,j_{-1}(x) \,P_{1}^2(x) \,P_{-2}(x), \,
\label{tsixb}
\ee
the first correction to this.
This induces a comparable expansion of the SW differential,
\be
\la \,=\, x { {\rm d} y \over y} \,=\,   \la_I  \,+\,  \la_{II} \,+...\,,
\label{tsixc}
\ee
where 
\be
\lambda_{I} \,=\,   
{ x\left( {A'\over A}-{B'\over {2B}}\right)
\over{\sqrt{1-{B\over A^2}}}}
{\rm d}x, 
\label{tsixd}
\ee
and \cite{product}
\be
\lambda_{II}   
\ =\  -\ {L^2\over 2} 
\left[  { j_{ 1}(x)  \,   P_{0}(x)  \, P_{ 2}(x)  \over   P_{ 1}^2 (x)  }
  \ + \ { j_{-1}(x)  \,   P_{0}(x)  \, P_{-2}(x)  \over   P_{-1}^2 (x)   }
 \right] {\rm d}x.
\label{correct}
\ee

Following the Seiberg-Witten approach,
we will now use the curve (\ref{seventeen}) 
together with the SW differential (\ref{tsixc})-(\ref{correct})
to compute the renormalized order parameters $a_k$ and their duals $a_{D,k}$
using eq.~(\ref{three}).
Our goal will be to choose $P_n(x)$ and $j_n(x)$
so that $a_{D,k}$ computed from the curve
agrees with eq.~(\ref{elevenc}).

The hyperelliptic curve (\ref{eighteen}) has two sheets 
connected by $N$ branch cuts 
extending from $x_k^-$ to $x_k^+$ 
and centered about $x=e_k$, the zeros of $P_0(x)$.
We choose the canonical homology basis as follows:
the cycle $A_k$ is a simple contour 
enclosing the branch cut centered about $e_k$;
the cycle $B_k$ goes 
from $x_1^{-}$ to $x_k^{-}$ on one sheet and 
from $x_k^{-}$ to $x_1^{-}$ on the other.
The order parameter 
$2\pi i a_{k} = \oint_{A_k} \la$
is calculated as in ref. \cite{DHokerKricheverPhong1}
\be
a_k =  e_k + O(L^2).
\ee
{}From Sec. 5 of ref. \cite{oneanti}, one of the contributions 
to the first approximation to the dual order parameter
$(2\pi i a_{D,k})_I = \oint_{B_k} \la_I$ is 
\be
{1\over 2}\int_{x_1^-}^{x_k^-} {\rm d}x\,{B\over A^2}
\,=\,2\,L^2\int_{x_1^-}^{x_k^-} {\rm d}x\,
{ j_0(x)  P_1(x)  P_{-1}(x)  \over P_0^2(x)  }.
\label{ttwo}
\ee
The integrand has second-order poles at $x=e_k$ 
from the factor $1/ P_0^2(x)$. 
The coefficients of these poles are chosen to be 
$S_k(a_k)$, in analogy with the first three entries of Table 1,
{\it i.e.},
\be
{4\,L^2\,j_0(x)\,P_{-1} (x)\,P_{1}(x)\over P_0^2(x)}
= L^2\,\sum_k {S_k (a_k) \over (x-e_k)^2}  + \cdots\,. 
\label{tthree}
\ee 
As a result, eq.~(\ref{ttwo}) will produce 
the term 
$
- \half L^2 \sum_{i\neq k}  S_i(a_i) /(a_k-a_i)  
$
in eq.~ (\ref{elevenc}). 
Equation (\ref{tthree}) is attained by setting
\be
j_0(x)\,{P_{-1}(x)P_{1}(x)\over P_0^2(x)}
= {1\over 4} S(x) + O(L^2),
\label{tfive}
\ee
where (from the last entry of Table 1)
\be
S(x) =    {4 \prod_{i=1}^N(x+a_i+2m_1)\, 
             \prod_{i=1}^N (x+a_i+2m_2)\,
             \prod_{j=1}^{N_f} (x+M_j)    \over 
             (x+m_1)^2\,(x+m_2)^2\, \prod_i (x-a_i)^2}  .
\ee
The correction to the dual order parameter 
\cite{product}
\be
(2\pi i a_{D,k})_{II}\,
=\,2\int_{x_1^-}^{x_k^-}\,\la_{II}\,
=\,-L^2\int_{x_1^-}^{x_k^-} {\rm d}x
\left[{j_{ 1}(x)    P_0(x)   P_{ 2}(x)    \over P_{ 1}^2 (x)   }\,+ \,
      {j_{-1}(x)    P_0(x)   P_{-2}(x)    \over P_{-1}^2 (x)   }\right],
\label{tseven}
\ee
evaluated using Sec. 5(b) of ref. \cite{oneanti}, 
gives rise to the term 
$ \quarter L^2 \sum_{\ell=1}^2 \sum^N_{i=1} {S_i(a_i)/ (a_k+a_i+2m_\ell)} $
in eq.~(\ref{elevenc})
if we choose
\be
j_1(x)\,{P_0(x)P_2(x)\over P_1^2(x)}\,=\,
{1\over 4}\,S(-x-2m_2) + O(L^2),
\label{teight}
\ee
and 
\be
j_{-1}(x)\,{P_{-2}(x) P_{0}(x) \over P_{-1}^2(x)}\,=
\,{1\over 4}\,S(-x-2m_1) + O(L^2).
\label{tnine}
\ee
The remaining terms of eq.~(\ref{elevenc}) arise from subleading (in $L$)
terms of the coefficient functions, discussed later in this section.

Observe from the right-hand side of (\ref{tfive}), 
(\ref{teight}) and (\ref{tnine}) that the ratio 
\be
j_n(x)\,{P_{n-1}(x)\,P_{n+1}(x)\over P_n^2(x)}
\ee
is invariant, 
up to a predictable reflection and shift in the argument of $S(x)$. 
This fact will be useful 
in understanding the M-theory interpretation of our results. 

Equations (\ref{tfive}), (\ref{teight}), and (\ref{tnine}) 
suffice to determine the leading order terms of $P_n(x)$ and $j_n(x)$.
The general solution to these equations is
\be
P_{ 2} (x)  &=& F(x) \, G(x)^{ 2} \, (x+m_2)^{-6} \,(x+2m_2-m_1)^{-2} \,
                \prod_{i=1}^N(x-a_i+2m_2-2m_1) + O(L^2),	\nn\\
P_{ 1} (x)  &=& F(x) \, G(x)  \,    (x+m_2)^{-2} \,
                (-1)^N \prod_{i=1}^N(x+a_i+2m_2) + O(L^2),		\nn\\ 
P_{ 0} (x)  &=& F(x) \,
                \prod_{i=1}^N(x-a_i) + O(L^2),  		\nn\\
P_{-1} (x)  &=& F(x) \, G(x)^{-1} \,     (x+m_1)^{-2} \,
                (-1)^N \prod_{i=1}^N(x+a_i+2m_1) + O(L^2),		 \nn\\
P_{-2} (x)  &=& F(x) \, G(x)^{-2}  \,   (x+m_1)^{-6} \,(x+2m_1-m_2)^{-2} \,
                \prod_{i=1}^N(x-a_i+2m_1-2m_2) +O(L^2),		\nn\\
j_{ 1} (x)  &=&(-1)^{N_f}\prod_{j=1}^{N_f}\,(x+2m_2-M_j),  		\nn\\
j_{ 0} (x)  &=&\prod_{j=1}^{N_f}\,(x+M_j),				\nn\\
j_{-1} (x)  &=&(-1)^{N_f}\prod_{j=1}^{N_f}\,(x+2m_1-M_j),	
\label{coeff}
\ee
where $F(x)$ and $G(x)$ are arbitrary functions.
The function $F(x)$ can be simply factored out of the curve,
and $G(x)$ eliminated by the change of variables $t \to t/G(x) $.

A check of these coefficient functions is obtained
if one of the antisymmetric hypermultiplets is removed from the spectrum
by letting its mass go to infinity.
One may verify that in the limit $m_2 \to \infty$, 
the quartic curve given by (\ref{seventeen}) and (\ref{coeff}) reduces, 
after the redefinition
\be
L^2 \lra  m_2^2 (-2m_2)^{-N} L^2,
\label{changeofvar}
\ee
to the cubic curve \cite{LandsteinerLopezLowe,nonhyper} 
for a single hypermultiplet in the antisymmetric representation
and $N_f$ hypermultiplets in the defining representation,
for leading terms of the coefficient functions.
The same result holds in the $m_1 \to \infty$ limit, 
in light of the involution (\ref{fourteen}).

Consideration of the one-instanton contribution to the prepotential 
also allows us to place some contraints on (but not uniquely determine) 
the subleading (in $L$) terms of $P_n(x)$.
We postulate that the coefficient functions 
in eq.~(\ref{twelve}) have subleading terms of the form
\be
P_{ 1}(x)
& = & (x+m_2)^{-2} \left[ (-)^N  \prod_{i=1}^N(x+a_i+2m_2) 
 \,+\, L^2  Q(-x-2m_2) \, + \, O(L^4) \right],  \nn \\
P_0(x)
& = &
\prod_{i=1}^N(x-a_i) \,+\, L^2  Q(x) + O(L^4),  \nn \\
P_{-1}(x)
& = & (x+m_1)^{-2} \left[ (-1)^N \prod_{i=1}^N(x+a_i+2m_1)  
\,+\, L^2  Q(-x-2m_1) + O(L^4) \right],  
\label{sublead}
\ee
with
\be
Q(x) &=&   
  {3\Aone \over   (x+m_1)^2}	  \,+\,  {\Bone \over   (x+m_1)}      \,+\,   
  {3\Atwo \over   (x+m_2)^2} 	  \,+\,  {\Btwo \over   (x+m_2)}      \,+\,
  {3\Athree \over (x+2m_2-m_1)^2} \,+\,  \nn\\[.1in]
&&{\Bthree \over (x+2m_2-m_1)} \,+\, 
  {3\Afour \over  (x+2m_1-m_2)^2} \,+\,  {\Bfour \over  (x+2m_1-m_2)} \,+\, 
  {3\Afive \over  (x+3m_1-2m_2)^2}\,+\,  \nn\\[.1in]
&&{\Bfive \over  (x+3m_1-2m_2)}\,+\,   
  {3\Asix \over   (x+3m_2-2m_1)^2}\,+\,  {\Bsix \over   (x+3m_2-2m_1)}\,+\,
\cdots \label{Qdefin}
\ee
The expression (\ref{Qdefin}) for $Q(x)$
is motivated by the results of sec.~4 of this paper,
in which an infinite number of orientifold sixplanes are required
in order to satisfy the reflection symmetries of the curve.
Eq.~(\ref{Qdefin}) reduces to 
\be
Q(x) \lra  {3\Aone \over   (x+m_1)^2}	  \,+\,  {\Bone \over   (x+m_1)}   
\ee
in the limit  $m_2\to \infty$,
in agreement with the form of curve for one antisymmetric 
and $N_f$ defining hypermultiplets \cite{LandsteinerLopezLowe,nonhyper}.

The constants 
$\Aone$, $\Bone, \ldots$ in eq.~(\ref{Qdefin})
are constrained by:

\noindent(a) the involution symmetry (\ref{fourteen}),

\noindent(b) the absence of 
$\log (a_k+m_1)$, $\log (a_k+2m_2-m_1)$, etc. terms  in $a_{D,k}$,

\noindent(c) the absence of poles in
$(a_k+3m_2-2m_1)$, $(a_k+4m_2-3m_1)$, etc. in $a_{D,k}$,

\noindent(d) agreement of the coefficients of the simple poles
at $(a_k+m_1)$, $(a_k+2m_2-m_1)$, $(a_k+m_2)$, and $(a_k+2m_1-m_2)$
in $a_{D,k}$ calculated from the curve 
with those in the expression (\ref{elevenc}), and

\noindent(e) the correct limiting behavior of the subleading terms as
$m_1 \to \infty$ and $m_2 \to \infty$.

These constraints give rise to a set of recursion relations 
that determine all but two of the constants.
We are unaware of any additional field-theoretic constraints 
that would allow us to fix the subleading terms of the curve uniquely.

We mention that if one arbitrarily truncates the expression $Q(x)$ 
to a {\it finite} number of terms,
the constants are uniquely determined by the constraints.
The resulting curve, however, 
has an unphysical singular limit as $m_2 \to m_1$.
We therefore conclude that such a truncation is inconsistent.

\bigskip
\noindent{\bf 4. M-theory picture}
\renewcommand{\theequation}{4.\arabic{equation}}
\setcounter{equation}{0}

In section 3,
we showed that the quartic SW curve 
\be
& & 	L^4 \, j_{1}(x) P_{ 2} (x)  			\, t^2 
\,+\,  	L   \,          P_{ 1} (x) 			\, t
\ +\          		P_{ 0} (x)          		\  	\cr 
& & +\, \ 	L      j_0(x)   P_{-1} (x)         	\, t^{-1}
\ +\   	L^4 \, j_0^2(x) j_{-1} (x)   P_{-2} (x)   	\, t^{-2}  
\,=\,0,
\label{twelverepeat}
\ee 
with coefficient functions (given to leading order in $L$) given 
by eqs.~(\ref{coeff}) ff,  as
\be
P_{ 2} (x)  &=& (x+m_2)^{-6} \,(x+2m_2-m_1)^{-2} \,
                \prod_{i=1}^N(x-a_i+2m_2-2m_1) + O(L^2),	\nn\\
P_{ 1} (x)  &=& (x+m_2)^{-2} \,
            \left[ (-)^N  \prod_{i=1}^N(x+a_i+2m_2) 
               \,+\,  O(L^2) \right],    \nn \\
P_{ 0} (x)  &=& \prod_{i=1}^N(x-a_i) \,+\,  O(L^2),  \nn \\
P_{-1} (x)  &=& (x+m_1)^{-2} \,
            \left[ (-1)^N \prod_{i=1}^N(x+a_i+2m_1)  
			\,+\, O(L^2) \right],  \nn \\
P_{-2} (x)  &=& (x+m_1)^{-6} \,(x+2m_1-m_2)^{-2} \,
                \prod_{i=1}^N(x-a_i+2m_1-2m_2) +O(L^2),		\nn\\
j_{ 1} (x)  &=& (-1)^{N_f}\prod_{j=1}^{N_f}\,(x+2m_2-M_j),  		\nn\\
j_{ 0} (x)  &=& \prod_{j=1}^{N_f}\,(x+M_j),				\nn\\
j_{-1} (x)  &=& (-1)^{N_f}\prod_{j=1}^{N_f}\,(x+2m_1-M_j),		
\label{coeffrepeat}
\ee
gives rise to the one-instanton prepotential (\ref{eleven}) 
postulated in section 2
for SU($N$) gauge theory with 
two antisymmetric hypermultiplets and $N_f$ defining hypermultiplets.
In this section,
we interpret this curve in terms of M-theory 
\cite{Mtheory}-\cite{LandsteinerLopezLowe}.
We then  present evidence that the curve (\ref{twelverepeat}),
though sufficient to generate $\cF_{\oneinst}$,
is only an effective curve,
and is in fact embedded in an infinite power series in $t$.

We begin by attempting to associate an M-theory picture\footnote{We 
describe the brane structure in terms of type IIA
string theory, which then lifts to M-theory~\cite{Witten}.}
with the quartic curve (\ref{twelverepeat}).
Such a picture involves four parallel NS 5-branes, 
with each adjacent pair connected by $N$ parallel D4-branes (see figure 1). 
The factors 
$ \prod_{i=1}^N(x+a_i+2m_2) $,
$ \prod_{i=1}^N(x-a_i) $, and
$ \prod_{i=1}^N(x+a_i+2m_1) $,
in $P_{1}$, $P_0$, and $P_{-1}$, respectively,
determine the positions of the connecting D4-branes.
There are also D6-branes between the NS 5-branes,
which correspond to 
the factors of $j_n(x)$ in eq.~(\ref{twelverepeat})
representing the hypermultiplets in the defining representation \cite{Witten}.

As noted in the previous section,
the quartic curve (\ref{twelverepeat}) 
reduces in the limit $m_2 \to \infty$ to a cubic curve 
describing SU$(N)$ gauge theory with one antisymmetric hypermultiplet
of mass $2m_1$.
In the M-theory picture (fig. 1),
in this limit the D4-branes dependent on $m_2$ slide off to infinity,
and the NS 5-brane denoted by $\NSone$ becomes disconnected,
leaving three parallel NS 5-branes connected by $N$ parallel D4-branes.
(In the limit $m_1 \to \infty$, 
the NS 5-brane $\NSfour$ becomes disconnected instead.)

Recall that the M-theory picture for a single antisymmetric hypermultiplet 
involves a negative charge orientifold sixplane (O$6^-$) 
on the central of three parallel NS 5-branes \cite{LandsteinerLopezLowe}.  
This pair of  O$6^{-}$ planes must also be present 
{\it before} the limits $m_1 \to \infty$ and $m_2 \to \infty$ are taken,
and we indicate their positions in fig. 1
by $\otimes$'s on $\NStwo$ and $\NSthree$.
The factors (and exponents) of $(x+m_1)$ and $(x+m_2)$
in the coefficient functions (\ref{coeffrepeat}) 
are exactly those expected 
for O$6^-$ planes in these locations.
To see this, recall that the presence of an O$6^-$ plane at
$x = -m_1$ implies that the geometry far from the orientifold
is represented by the complex manifold \cite{LandsteinerLopezLowe}
\be
t \hat{t} = {L^2 \, j_0(x)\,  j_{-1}(x) \over (x+m_1)^4}
\ee
and that the curve should be invariant under the orientifold projection
\be
x \lra -x-2m_1
\hs{.3in} ; \hs{.3in}
t \llra \hat{t}.
\label{projection}
\ee
Imposing this invariance on the {\it last four terms} of the curve 
(\ref{twelverepeat}) 
(those that remain when $m_2 \to \infty$)
yields the relations
$ P_{-2} (x) = (x+m_1)^{-6} P_{ 1} (-x-2m_1)$
and
$ P_{-1} (x) = (x+m_1)^{-2} P_{ 0} (-x-2m_1)$,
in agreement with (\ref{coeffrepeat}).
A similar story holds for the factors of $(x+m_2)$.
The fact that the full quartic curve (\ref{twelverepeat}) 
is not invariant under the projection (\ref{projection})
is the first indication that this curve is incomplete.

O$6^-$ planes represent reflection symmetries in the M-theory picture,
and the reflections of the two O$6^{-}$ planes on $\NStwo$ and $\NSthree$
generate an infinite number of parallel NS 5-branes, 
with an O$6^{-}$ plane on each of them.
The factors of $(x+2m_2-m_1)$ and $(x+2m_1-m_2)$
in the coefficient functions (\ref{coeffrepeat}) 
exactly correspond to O$6^-$ planes on $\NSone$ and $\NSfour$,
and the factors of 
$\prod_{i=1}^N(x-a_i+2m_2-2m_1) $
and 
$\prod_{i=1}^N(x-a_i+2m_1-2m_2) $
in $P_{ 2} $ and $ P_{-2} $, respectively,
represent D4-branes connecting  
$\NSone$ and $\NSfour$ to the rest
of the infinite chain of NS 5-branes.

The necessary presence of additional NS 5-branes may equivalently be seen
by requiring the SW curve to possess the involution symmetries implied by 
the O$6^-$ planes at $x=-m_1$ and $x=-m_2$.
As we saw above,
the last four terms of the curve (\ref{twelverepeat}) 
are invariant under the involution
\be
x \lra -x-2m_1
\hs{.3in} ; \hs{.3in}
t \lra {L^2 \, j_0(x)\,  j_{-1}(x) \over (x+m_1)^4 \, t}, 
\label{firstinvolution}
\ee
but invariance of the full curve requires the presence of a $t^{-3}$ term.
Similarly,
the first four terms of the curve (\ref{twelverepeat}) are invariant
under the involution
\be
x \lra -x-2m_2
\hs{.3in} ; \hs{.3in}
t \lra {(x+m_2)^4 \over L^2 \, t}, 
\label{secondinvolution}
\ee
but invariance of the full curve requires the presence of a $t^{3}$ term.
Only a curve of infinite order can be 
simultaneously invariant under both involutions.

Figure 2 represents an expanded view of the brane configuration,
involving six parallel NS 5-branes. 
The positions of the D4-branes, D6-branes, and O$6^-$ planes
are completely dictated by reflection symmetries.
The sextic curve associated with this truncation 
of the infinite chain of branes is
\be
& & 	L^9 \, j_{1}^2(x) j_{ 2}(x) 	P_{ 3} (x)  	\, t^3 
\,+\, 	L^4 \, j_{1}(x) 		P_{ 2} (x)	\, t^2 
\,+\,  	L   \,          		P_{ 1} (x)	\, t
\ +\          				P_{ 0} (x)  
\ +\, 	L      j_0(x)  			P_{-1} (x)  	\, t^{-1} \nn\\
& & \ +\   	L^4 \, j_0^2(x) j_{-1}(x)   	P_{-2} (x)   	\, t^{-2}  
\ +\   	L^9 \, j_0^3(x) j_{-1}^2(x) j_{-2}(x) P_{-3} (x)\, t^{-3}
\,=\,0, 
\label{sextic}
\ee 
and would be required to compute the prepotential to three-instanton accuracy.
The involution symmetries (\ref{firstinvolution}) and (\ref{secondinvolution}) 
determine the new coefficients (to leading order in $L$) to be
\be
P_{ 3} (x)  &=& (x+m_2)^{-10} \, (x+2m_2-m_1)^{-6} \,(x+3m_2-2m_1)^{-2} \,
                (-1)^N \prod_{i=1}^N(x+a_i+4m_2-2m_1),		\nn\\
P_{-3} (x)  &=& (x+m_1)^{-10} \, (x+2m_1-m_2)^{-6} \,(x+3m_1-2m_2)^{-2} \,
                (-1)^N \prod_{i=1}^N(x+a_i+4m_1-2m_2),		\nn\\
j_{ 2} (x)  &=& \prod_{j=1}^{N_f}\,(x+2m_2-2m_1+M_j),  		\nn\\
j_{-2} (x)  &=& \prod_{j=1}^{N_f}\,(x+2m_1-2m_2+M_j).		
\label{newcoeff}
\ee
Observe from eqs.~(\ref{coeffrepeat}) and (\ref{newcoeff}) that the ratio
\be
j_n(x)\,{P_{n-1}(x) P_{n+1}(x)\over P_n^2(x)}
\label{again}
\ee
is identical to (\ref{tfive}) 
up to a  predictable reflection and shift in the argument of $S(x)$. 
One may take this as a general principle, 
which implies that one may take {\it any} pair
of adjacent parallel 5-branes to define a hyperelliptic approximation for
the instanton expansion. 
This observation leads to results equivalent to
the successive imposition of involution symmetries  such as 
(\ref{firstinvolution}) and (\ref{secondinvolution}) for all possible 
embedded quartic curves.

Figure 3 shows that in the $m_2\to \infty$ limit,
the M-theory picture of fig.~2
reduces to that of one antisymmetric representation 
with an O$6^-$ plane on NS 5-brane $\NSthree$.
Had we taken $m_1\to \infty$ instead,
the analogue to figure 3 would have involved 5-branes 
$\NSone$, $\NStwo$, and  $\NSthree$
connected by D4 branes with an O$6^-$ plane on NS 5-brane $\NStwo$. 

The full curve  describing SU($N$) gauge theory with two antisymmetric 
and $N_f$ defining hypermultiplets, 
in which the quartic (\ref{twelverepeat}) 
and sextic (\ref{sextic}) approximations are embedded, is
\be
\sum_{n=1}^\infty  L^{n^2} \prod_{s=1}^{n-1} j_s^{n-s}(x) P_n(x) t^n +
P_0(x) +
\sum_{n=1}^\infty  L^{n^2} j_0^n(x) 
\,\prod_{s=1}^{n-1} j_{-s}^{n-s}(x) P_{-n}(x) t^{-n}=0.
\label{fullcurve}
\ee
The involution symmetries (\ref{firstinvolution}) and (\ref{secondinvolution}) 
imply the following recursion relations
\be
P_{-n-1} (x) &=& (x+m_1)^{-4n-2} P_{ n} (-x-2m_1), \nn\\
P_{n+1}  (x) &=& (x+m_2)^{-4n-2} P_{-n} (-x-2m_2), \nn\\
j_{-n-1} (x) &=& j_{ n} (-x-2m_1), 		\nn\\
j_{n+1}  (x) &=& j_{-n} (-x-2m_2), 
\label{recursion}
\ee
which fully determine the coefficient functions of the curve
(including subleading terms)
in terms of $P_0(x)$.
Note that the coefficients determined from
the recursion relations  (\ref{recursion}) 
imply that the ratio (\ref{again}) is invariant, 
up to a predictable reflection and shift in the argument of $S(x)$.

\bigskip
\noindent{\bf 5. Summing the Series}
\renewcommand{\theequation}{5.\arabic{equation}}
\setcounter{equation}{0}

In this section we sum the infinite series (\ref{fullcurve}) 
for certain special cases, 
obtaining results in terms of theta functions.  
The solution to the recursion relations (\ref{recursion}) 
allows one to write (\ref{fullcurve}) explicitly,
keeping the leading terms only, 
\be
&&
\sum_{n=-\infty}^{n=\infty}\, L^{n^2}t^n (-1)^{Nn} 
j_0^{{1\over 2}(|n|-n)} (x) 
\prod_{s=1}^{|n|-1} 
j_{ns / |n|}^{|n|-s} (x) 
\nn\\
&\times& 
\prod_{i=1}^N\,\left[x-(-1)^na_i-n\,\Delta
+\half (1-(-1)^n) (m_1+m_2)  \right] \nn\\
&\times& 
\prod_{p=0}^{|n|-1} \,\left[
x+m_1-n\,\Delta + \left( {{n-|n|} \over  2|n|}\right)\Delta
+{n\over |n|}p\Delta
\right]^{-(4p+2)}=0, 
\label{ione}
\ee
where $\Delta=m_1-m_2$, and
\be
j_0(x)&=& \prod_{j=1}^{N_f}(x+M_j), \nn\\
j_{ns / |n|} (x) 
&=&
(-1)^{s  N_f}
\prod_{j=1}^{N_f}\left[
x-{n\over |n|}s\,\Delta
+(-1)^sM_j
+\half (1-(-1)^s) (m_1+m_2)
\right].
\label{itwo}
\ee

Rather than attempting to sum (\ref{ione}) for the general case, 
let us consider the special case
$m_1=m_2=m$ ({\it i.e.}, $\Delta=0$),  
for which the curve simplifies to 
\be
&&\sum_{n=-\infty}^{n=\infty}\,{ L^{n^2}t^n\over (x+m)^{2n^2}}\, 
j_0^{{1\over 2}(|n|-n)} (x) \, 
\prod_{s=1}^{|n|-1} 
j_{ns / |n|}^{|n|-s} (x) 
\nn\\
&\times& 
\left( (-1)^{Nn} \prod_{i=1}^N\,\left[x-(-1)^na_i+ (1-(-1)^n)m\right]
\, +\, O(L^2) \right)=0, 
\label{ithree}
\ee
where
\be
\prod_{s=1}^{|n|-1} j_{ns / |n|}^{|n|-s} (x) 
&=& 
j_{0}^{|n|(|n|-2)/4} (x) \,j_{1}^{n^2/ 4} (x) 
\hs{.7in}{\rm for}\,\,\,n\,\,\,{\rm even}, \nn\\
&=& 
j_{0}^{(|n|-1)^2/4}(x)\,j_{1}^{(n^2-1)/4} (x)
\hs{.5in}{\rm for}\,\,\,n\,\,\,{\rm odd},
\label{ifour}
\ee
with
\be
j_{0}(x)&=&\prod_{j=1}^{N_f}(x+M_j), 
\nn\\
j_{1}(x)&=&(-1)^{N_f}\,\prod_{j=1}^{N_f}(x+2m-M_j).
\label{isix}
\ee
Defining
\be
H_0(x)&=&\prod_{i=1}^N(x-a_i) \,\,+O(L^2),\nn\\
H_1(x)&=&(-1)^N \,\prod_{i=1}^N(x+a_i+2m)\,\,+O(L^2),
\label{ieight}
\ee
we rewrite the curve (\ref{ithree}) as
\be
&&
H_0(x)\sum_{n\,{\rm even}}
\,\left[L\,j_{0}^{1/4}(x) \,j_{1}^{1/4}(x) \over (x+m)^2\right]^{n^2}\, 
\left(t\,j_{0}^{-1/2}(x) \right)^n \nn\\
&+&
H_1(x)\,\left({j_{0}(x)\over j_{1}(x)}\right)^{1/4}\,\,\sum_{n\,{\rm odd}}
\left[L\,j_{0}^{1/4}(x)\, j_{1}^{1/4}(x)\over (x+m)^2\right]^{n^2}\,
\left(t\,j_{0}^{-1/2}(x) \right)^n =0,
\label{inine}
\ee
which may be expressed as
\be
H_0(x)\,\sum_{n=-\infty}^{\infty}\,q(x)^{n^2}\,\tilde t^n
\,+\,
\left({j_{0}(x)\over j_{1}(x)}\right)^{1/4}\,H_1(x)\,
\sum_{n=-\infty}^{\infty} \,q(x)^{(n+\half)^2}\,
\tilde t^{(n+\half)}\,=\,0,
\label{itwelve}
\ee
with
\be
q(x)\,=\,{L^4\,j_{0}(x)\,j_{1}(x)\over (x+m)^8}
\,\,\,\,\,\,{\rm and}\,\,\,\,\,\,
\tilde t\,=\,t^2\,j^{-1}_{0}(x).
\label{iten}
\ee
The curve (\ref{itwelve}) may now be recast in terms of theta functions as 
\be
H_0(x)\,\theta_3(s|q(x))\,
+
\left({j_{0}(x)\over j_{1}(x)}\right)^{1/4}
\,H_1(x)\,\theta_2(s|q(x))\,=\,0,
\label{ififteen}
\ee
where $ \tilde t=\e^{2\pi i s}$ and 
\be
\theta_3 \,(s|q)
&=&\sum_{n=-\infty}^{\infty}\,q^{n^2}\,\e^{2\pi i ns},\nn\\
\theta_2 \,(s|q)
&=&\sum_{n=-\infty}^{\infty}\,q^{(n+{1\over 2})^2}\,\e^{2\pi i (n+{1\over 2})s}.
\label{ifourteen}
\ee
Since the theta functions (\ref{ifourteen}) are only defined for $|q(x)|<1$,
however,
we observe from eq.~(\ref{iten}) 
that the series (\ref{itwelve}) 
is not well-defined for $x \to -m$, nor for large $L$.
This may be an indication that eq.~(\ref{ithree}) is inconsistent 
with the $O(L^2)$ subleading terms omitted.
It is possible that these subleading terms sum up to allow a 
continuation to regions where $|q(x)|>1$.
As we saw in sec.~3, however,
such subleading terms are not uniquely determined by 
the one-instanton prepotential.
Nor do we have a prediction of these subleading terms from M-theory.

Uranga \cite{Uranga} considers the scale invariant case of SU$(N)$ with 
two antisymmetric hypermultiplets and four defining hypermultiplets
This corresponds to an elliptic model \cite{DonagiWitten}
with a brane configuration of two NS 5-branes, 
with an O$6^{-}$ orientifold on each 
and two sets of $N$ D4-branes connecting the two NS 5-branes pair-wise
around a circle $S^1$ in the $t$-direction 
(or the $x_6$-direction in the notation of ref.~\cite{Witten}), 
together with $N_f=4\,$ D6-branes. 
Our results for $N_f\leq 3$ may be regarded as a
decompactification of Uranga's elliptic model, 
where $t$ is now the covering space of the $S^1$ of the elliptic model.
(For $\Delta=0$, this is an untwisted elliptic model.) 
This interpretation is supported by the fact that 
the terms of the series (\ref{ithree}) sum up to form theta functions.

Note also the work of Yokono \cite{Yokono}, 
who considered softly broken ${\cal N}=4\, $ 
USp$(2N_c)$ and SO$(N_c)$ gauge  theory. 
He finds a brane picture analogous to ours 
(see for example his figs. 2 and 3). 
An infinite number of parallel 5-branes and O$4^{-}$-planes 
result from the decompactification of an elliptic model.

We return to the curve (\ref{ione}), 
taking $N_f=0$ for simplicity. 
It is straightforward to expand the curve 
in powers of $\Delta=m_1-m_2$, 
and express the result in terms of theta functions. 
To first order in $\Delta$, 
keeping only the leading terms of $P_n(x)$,
we find
\be
&&\left(1-{2\Delta \over {3(x+m_1)}}
\left[ 8(\tilde t {\pr\over {\pr \tilde t}})\,(q(x){\pr\over {\pr q(x)}})
-6 q(x) {\pr\over {\pr q(x)}}+ \tilde t {\pr\over {\pr \tilde t}} \right]
\right)\,
\left[ H_0(x)\theta_3(s|q(x))+ H_1(x)\theta_2(s|q(x))\right]\nn\\
&& + 2\Delta[{\pr \over {\pr z}} H_0(x+z)]_{z=0}\,
\,(\tilde t {\pr\over {\pr \tilde t}})\, \theta_3(s|q(x))
+2\Delta[{\pr \over {\pr z}} H_1(x+z)]_{z=0}\,\,(\tilde t 
{\pr\over {\pr \tilde t}}-{1\over 2})\,\theta_2(s|q(x))\,=\,0\nn\\
\label{inineteen}
\ee
where 
eqs.~(\ref{ieight}), (\ref{iten}), and (\ref{ifourteen}) have been used, 
and $ \tilde t=\e^{2\pi i s}$ as before.
Again this is valid only for $|q(x)|<1$.
The derivation of (\ref{ififteen})
or the expected elliptic model from an integrable model
remains a challenging problem for future work.

\bigskip
\noindent{\bf 6. Summary}
\renewcommand{\theequation}{6.\arabic{equation}}
\setcounter{equation}{0}

{}From our previous work, 
we are able to exhibit sufficient universality 
in the form of $\cF_{\oneinst}$ for SU$(N)$
with matter hypermultiplets 
to present an extremely plausible form (\ref{eleven}) for $\cF_{\oneinst}$ for 
two antisymmetric and $0\leq N_f\leq 3$ defining hypermultiplets. 
Using methods developed in refs.~\cite{oneanti}--\cite{product}
for a systematic instanton expansion 
based on a perturbative expansion beginning 
with a hyperelliptic approximation to a SW curve, 
we were able to ``reverse engineer"
a quartic curve which reproduces $\cF_{\oneinst}$. 
The leading order terms in $L$ are unique,
and there are strong constraints on the subleading terms.
When the mass of either of the antisymmetric hypermultiplets
goes to infinity,
the curve reduces to that for 
one antisymmetric and $N_f$ defining hypermultiplets. 

The quartic curve constructed in this way led us to 
an M-theory picture containing four NS 5-branes, 
connected by D4-branes.
However, since there are also O$6^{-}$ planes on 
each of the parallel 5-branes,
we were forced to consider an infinite chain of 
5-branes and O$6^-$ planes.
A finite subset of $2d$ of these 5-branes yields
an effective curve of order $2d$,
which is necessary to compute the prepotential 
to $\half d(d-1)$-instanton accuracy.
Without requiring consistency with M-theory,
one could have stopped with the quartic curve of sec. 3, 
if the only input were $\cF_{\oneinst}$.  
A computation of $\cF_{\threeinst}$ from an underlying Lagrangian, 
which could be compared with the 3-instanton prediction 
of the sextic curve\footnote{This prediction is somewhat ambiguous
because the subleading coefficients of the curve are not known exactly.} 
of sec. 4, 
would therefore provide support for the M-theory picture we have developed.

It is interesting that 
the SW curve and M-theory picture 
for SU$(N)$ gauge theory with two antisymmetric hypermultiplets 
differs so radically from 
that with only one antisymmetric hypermultiplet. 
It is not a ``trivial" extension of known results,
as we had originally anticipated, 
but is in fact much richer.
Nevertheless, our curve and M-theory picture 
reduce to that of ref.~\cite{LandsteinerLopezLowe}
in the large $m_1$ or $m_2$ limit. 

The summation of the infinite series representing the curve
allowed us to represent it in terms of theta functions in sec. 5. 
This suggests that our curve may be related to the 
decompactification of a scale-invariant elliptic model.
Uranga \cite{Uranga} has discussed an M-theory picture for SU$(N)$ 
with two antisymmetric and four defining hypermultiplets,
a scale-invariant case, but without specifying a curve. 
We speculate that, were the curve for this theory known,
sending the mass of one or more of the defining hypermultiplets
to infinity would be consistent with our analysis.  

Softly broken ${\cal N}=4\, $ SO$(N_c)$ and USp$(2N_c)$ 
gauge theories have been considered by Yokono \cite{Yokono}. 
His M-theory picture appears to be compatible with our analysis as well, 
in the sense that there is a decompactification with 
an infinite number of equally spaced parallel 5-branes, 
and an infinite number of orientifold planes. 

Finally, we remark that the observed universality of form for 
$\cF_{\oneinst}$ shown in Table 1 is still without a satisfactory 
derivation from first principles, 
particularly when the curve is not hyperelliptic. 
It is clear from this Table, however,
that the renormalized order parameters are 
the natural variables for this problem,
as emphasized in refs.~\cite{DHokerKricheverPhong1}
and \cite{oneanti}--\cite{product}.

\vs{4mm}

\noindent{\bf{Acknowledgement:}} We would like to thank 
Ansar Fayyazuddin, \"Ozg\"ur Sar{\i}o\~{g}lu, and Edward Witten
for valuable discussions. 
HJS wishes to thank the Physics Department of Harvard University 
for their continued hospitality, 
and to the CERN theory group for hospitality during summer 1998.
\eject

\begin{center}
\begin{picture}(810,295)(10,10)


\put(10,192){\line(1,0){5}}
\put(20,192){\line(1,0){5}}
\put(30,192){\line(1,0){5}}
\put(40,192){\line(1,0){5}}
\put(50,192){\line(1,0){5}}
\put(60,192){\line(1,0){5}}
\put(70,192){\line(1,0){5}}
\put(80,192){\line(1,0){5}}
\put(90,192){\line(1,0){5}}
\put(100,192){\line(1,0){5}}
\put(5,180){$(\!x\!-\!a_i\!+\!2m_2\!-\!2m_1\!)$}

\put(105,240){\line(1,0){2}}
\put(110,240){\line(1,0){5}}
\put(120,240){\line(1,0){5}}
\put(130,240){\line(1,0){5}}
\put(140,240){\line(1,0){5}}
\put(150,240){\line(1,0){5}}
\put(160,240){\line(1,0){5}}
\put(170,240){\line(1,0){5}}
\put(180,240){\line(1,0){5}}
\put(190,240){\line(1,0){5}}
\put(200,240){\line(1,0){5}}
\put(120,245){$(\!x\!+\!a_i\!+\!2m_2\!)$}

\put(205,117){\line(1,0){2}}
\put(210,117){\line(1,0){5}}
\put(220,117){\line(1,0){5}}
\put(230,117){\line(1,0){5}}
\put(240,117){\line(1,0){5}}
\put(250,117){\line(1,0){5}}
\put(260,117){\line(1,0){5}}
\put(270,117){\line(1,0){5}}
\put(280,117){\line(1,0){5}}
\put(290,117){\line(1,0){5}}
\put(300,117){\line(1,0){5}}
\put(235,105){$(\!x-\!a_i\!)$}

\put(305,155){\line(1,0){2}}
\put(310,155){\line(1,0){5}}
\put(320,155){\line(1,0){5}}
\put(330,155){\line(1,0){5}}
\put(340,155){\line(1,0){5}}
\put(350,155){\line(1,0){5}}
\put(360,155){\line(1,0){5}}
\put(370,155){\line(1,0){5}}
\put(380,155){\line(1,0){5}}
\put(390,155){\line(1,0){5}}
\put(400,155){\line(1,0){5}}
\put(320,160){$(\!x\!+\!a_i\!+\!2m_1\!)$}

\put(405,10){\line(1,0){2}}
\put(410,10){\line(1,0){5}}
\put(420,10){\line(1,0){5}}
\put(430,10){\line(1,0){5}}
\put(440,10){\line(1,0){5}}
\put(450,10){\line(1,0){5}}
\put(460,10){\line(1,0){5}}
\put(470,10){\line(1,0){5}}
\put(480,10){\line(1,0){5}}
\put(490,10){\line(1,0){5}}
\put(500,10){\line(1,0){5}}
\put(407,15){$(\!x\!-\!a_i\!+\!2m_1\!-\!2m_2\!)$}


\put(105,5){\line(0,1){255}}
\put(205,5){\line(0,1){255}}
\put(305,5){\line(0,1){255}}
\put(405,5){\line(0,1){255}}
\put(103,271){$\NSone$}
\put(203,271){$\NStwo$}
\put(303,271){$\NSthree$}
\put(403,271){$\NSfour$}


\put(101,215){$\otimes$}
\put(201,175){$\otimes$}
\put(301,135){$\otimes$}
\put(401,85){$\otimes$}
\put(31,220){$(\!x\!+\!2m_2\!-\!m_1\!)$}
\put(162,170){$(\!x\!+\!m_2\!)$}
\put(262,135){$(\!x\!+\!m_1\!)$}
\put(406,75){$(\!x\!+\!2m_1\!-\!m_2\!)$}

\put(151,185){\framebox(5,5){$\cdot$}}
\put(251,165){\framebox(5,5){$\cdot$}}
\put(351,105){\framebox(5,5){$\cdot$}}
\put(120,192){$(\!x\!+\!2m_2\!-\!M_j\!)$}
\put(235,172){$(\!x\!+\!M_j\!)$}
\put(320,96){$(\!x+\!2m_1\!-\!M_j\!)$}

\put(30,30){\vector(1,0){20}}
\put(30,30){\vector(0,1){20}}
\put(25,50){$x$}
\put(50,25){$t$}

\end{picture}
\end{center}
\label{figureone}
{\footnotesize{\bf Figure 1}: 
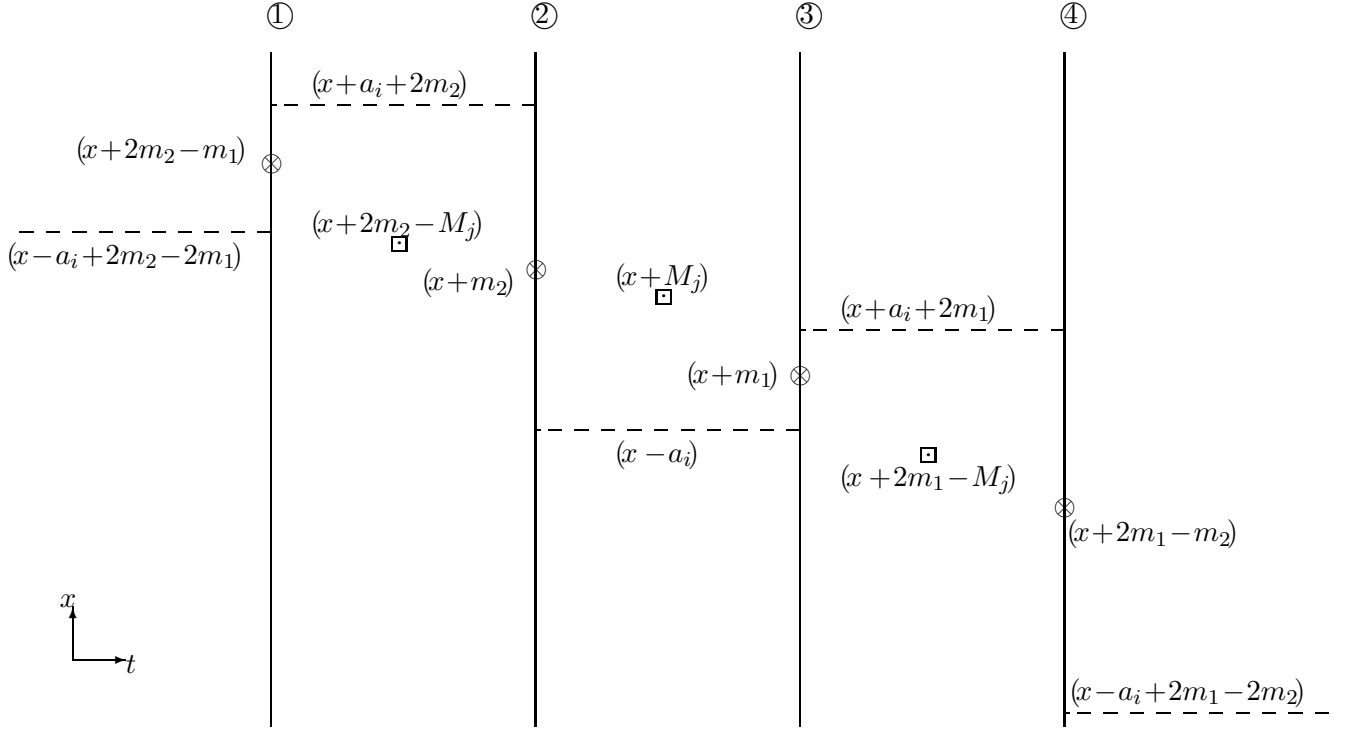
Four parallel, equally-spaced (in $t$) NS 5-branes are indicated 
by solid vertical lines, numbered $\NSone$ to  $\NSfour$ above. 
$N$ parallel D4-branes, linking each pair of adjacent 5-branes,
are indicated by horizontal dashed lines, with the position of 
the $i^{\rm th}$ 4-brane ($i=1\,\,{\rm to}\,\,N$) shown in the figure. 
D6-branes are indicated by~{\framebox(5,5){$\cdot$}},
and O$6^{-}$ planes by $\otimes$, 
with their positions in the $x$-direction shown. 
All elements in the figure respect the (multiple) 
mirror symmetries of the O$6^{-}$ planes. 
The reflection symmetries imply additional 4-branes to the left and right 
of 5-branes $\NSone$ and $\NSfour$, respectively, 
and an infinite number of parallel 5-branes, etc. 
Our $x$ corresponds to the variable $v=x_4+ix_5$ 
and $\ln t= -(x_6+ix_{10})/R$ of ref. \cite{Witten}.}
\eject


\begin{center}
\begin{picture}(810,400)(10,10)


\put(10,395){\line(1,0){5}}
\put(20,395){\line(1,0){5}}
\put(30,395){\line(1,0){5}}
\put(40,395){\line(1,0){5}}
\put(50,395){\line(1,0){5}}
\put(60,395){\line(1,0){5}}
\put(5,400){$\text{\scriptstyle{(\!x\!+\!a_i\!+\!4m_2-\!2m_1\!)}}$}

\put(65,240){\line(1,0){2}}
\put(70,240){\line(1,0){5}}
\put(80,240){\line(1,0){5}}
\put(90,240){\line(1,0){5}}
\put(100,240){\line(1,0){5}}
\put(110,240){\line(1,0){5}}
\put(120,240){\line(1,0){5}}
\put(130,240){\line(1,0){5}}
\put(72,245){$\text{\scriptstyle{(\!x\!-\!a_i\!+\!2m_2\!-\!2m_1\!)}}$}

\put(135,295){\line(1,0){2}}
\put(140,295){\line(1,0){5}}
\put(150,295){\line(1,0){5}}
\put(160,295){\line(1,0){5}}
\put(170,295){\line(1,0){5}}
\put(180,295){\line(1,0){5}}
\put(190,295){\line(1,0){5}}
\put(200,295){\line(1,0){5}}
\put(150,300){$\text{\scriptstyle{(\!x\!+\!a_i\!+\!2m_2\!)}}$}

\put(205,145){\line(1,0){2}}
\put(210,145){\line(1,0){5}}
\put(220,145){\line(1,0){5}}
\put(230,145){\line(1,0){5}}
\put(240,145){\line(1,0){5}}
\put(250,145){\line(1,0){5}}
\put(260,145){\line(1,0){5}}
\put(270,145){\line(1,0){5}}
\put(228,148){$\text{\scriptstyle{(\!x-\!a_i\!)}}$}

\put(275,190){\line(1,0){2}}
\put(280,190){\line(1,0){5}}
\put(290,190){\line(1,0){5}}
\put(300,190){\line(1,0){5}}
\put(310,190){\line(1,0){5}}
\put(320,190){\line(1,0){5}}
\put(330,190){\line(1,0){5}}
\put(340,190){\line(1,0){5}}
\put(290,193){$\text{\scriptstyle{(\!x\!+\!a_i\!+\!2m_1\!)}}$}

\put(345,45){\line(1,0){2}}
\put(350,45){\line(1,0){5}}
\put(360,45){\line(1,0){5}}
\put(370,45){\line(1,0){5}}
\put(380,45){\line(1,0){5}}
\put(390,45){\line(1,0){5}}
\put(400,45){\line(1,0){5}}
\put(410,45){\line(1,0){5}}
\put(350,48){$\text{\scriptstyle{(\!x\!-\!a_i\!+\!2m_1\!-\!2m_2\!)}}$}

\put(415,90){\line(1,0){2}}
\put(420,90){\line(1,0){5}}
\put(430,90){\line(1,0){5}}
\put(440,90){\line(1,0){5}}
\put(450,90){\line(1,0){5}}
\put(460,90){\line(1,0){5}}
\put(470,90){\line(1,0){5}}
\put(480,90){\line(1,0){5}}
\put(420,93){$\text{\scriptstyle{(\!x\!+\!a_i\!+\!4m_1\!-\!2m_2\!)}}$}


\put(65,5){\line(0,1){405}}
\put(135,5){\line(0,1){405}}
\put(205,5){\line(0,1){405}}
\put(275,5){\line(0,1){405}}
\put(345,5){\line(0,1){405}}
\put(415,5){\line(0,1){405}}
\put(63,421){$\NSzero$}
\put(133,421){$\NSone$}
\put(203,421){$\NStwo$}
\put(273,421){$\NSthree$}
\put(343,421){$\NSfour$}
\put(413,421){$\NSfive$}

\put(26,345){\framebox(5,5){$\cdot$}}
\put(96,285){\framebox(5,5){$\cdot$}}
\put(166,245){\framebox(5,5){$\cdot$}}
\put(236,185){\framebox(5,5){$\cdot$}}
\put(306,145){\framebox(5,5){$\cdot$}}
\put(376,85){\framebox(5,5){$\cdot$}}
\put(446,45){\framebox(5,5){$\cdot$}}
\put(1,337){$\text{\scriptstyle{(\!x\!+\!4m_2\!-\!2m_1\!-\!M_j\!)}}$}
\put(69,277){$\text{\scriptstyle{(\!x\!+\!2m_2\!-\!2m_1\!+\!M_j\!)}}$}
\put(149,237){$\text{\scriptstyle{(\!x\!+\!2m_2\!-\!M_j\!)}}$}
\put(230,177){$\text{\scriptstyle{(\!x\!+\!M_j\!)}}$}
\put(289,137){$\text{\scriptstyle{(\!x+\!2m_1\!-\!M_j\!)}}$}
\put(347,77){$\text{\scriptstyle{(\!x\!+\!2m_1\!-\!2m_2\!+\!M_j\!)}}$}
\put(420,37){$\text{\scriptstyle{(\!x\!+\!4m_1\!-\!2m_2\!-\!M_j\!)}}$}


\put(61,315){$\otimes$}
\put(131,265){$\otimes$}
\put(201,215){$\otimes$}
\put(271,165){$\otimes$}
\put(341,115){$\otimes$}
\put(411,65){$\otimes$}
\put(71,315){$\text{\scriptstyle{(\!x\!+\!3m_2\!-\!2m_1\!)}}$}
\put(141,265){$\text{\scriptstyle{(\!x\!+\!2m_2\!-\!m_1\!)}}$}
\put(211,215){$\text{\scriptstyle{(\!x\!+\!m_2\!)}}$}
\put(281,165){$\text{\scriptstyle{(\!x\!+\!m_1\!)}}$}
\put(351,115){$\text{\scriptstyle{(\!x\!+\!2m_1\!-\!m_2\!)}}$}
\put(421,65){$\text{\scriptstyle{(\!x\!+\!3m_1\!-\!2m_2\!)}}$}

\put(30,30){\vector(1,0){20}}
\put(30,30){\vector(0,1){20}}
\put(25,50){$x$}
\put(50,25){$t$}


\end{picture}
\end{center}
\label{figuretwo}
{\footnotesize{\bf Figure 2}: 
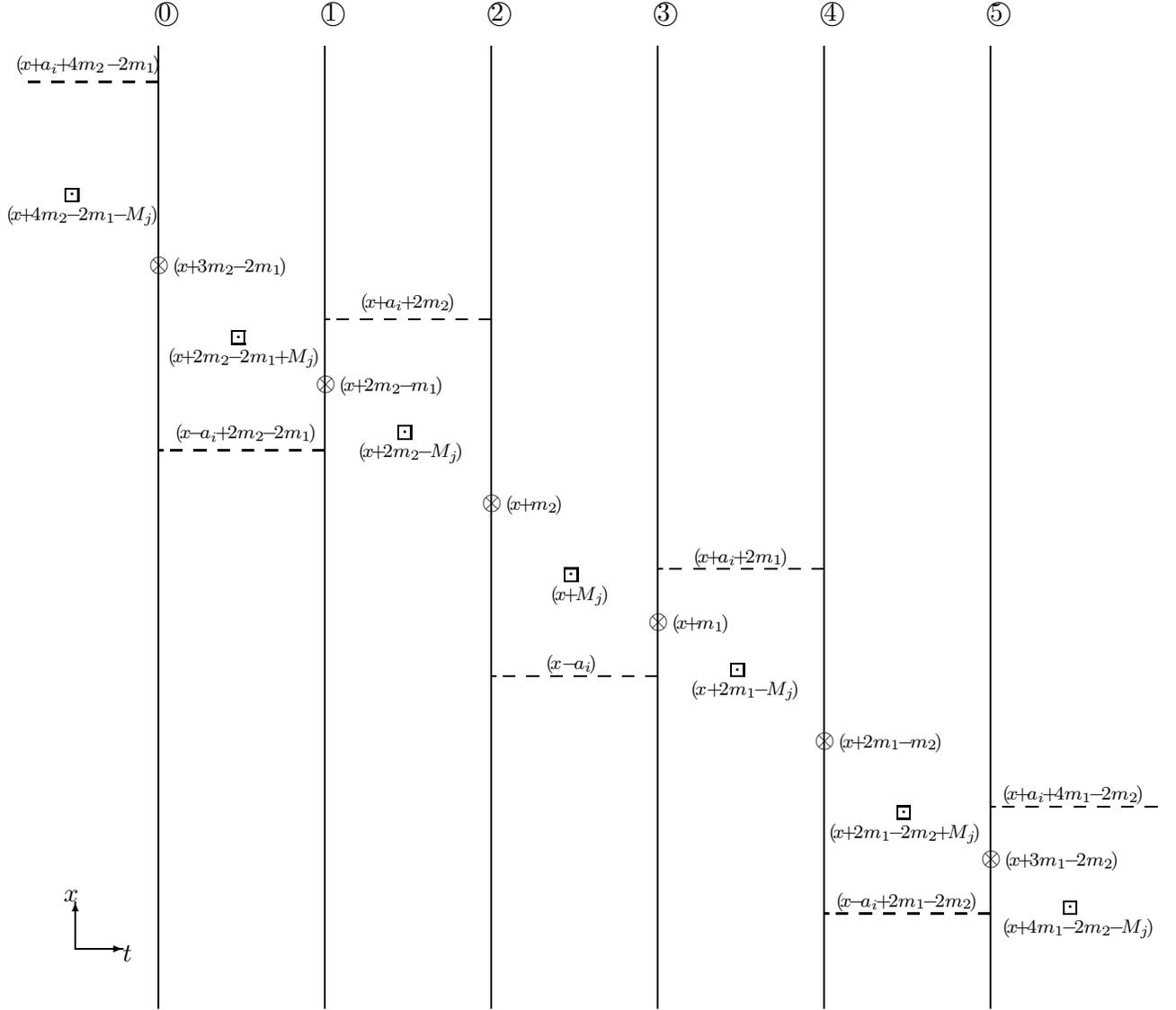
An expanded version of figure 1 with six NS 5-branes.}

\eject

\begin{center}
\begin{picture}(810,295)(10,10)


\put(195,117){\line(1,0){2}}
\put(200,117){\line(1,0){5}}
\put(210,117){\line(1,0){5}}
\put(220,117){\line(1,0){5}}
\put(230,117){\line(1,0){5}}
\put(240,117){\line(1,0){5}}
\put(250,117){\line(1,0){5}}
\put(260,117){\line(1,0){5}}
\put(268,117){\line(1,0){2}}
\put(210,105){$(\!x-\!a_i\!)$}

\put(270,155){\line(1,0){2}}
\put(275,155){\line(1,0){5}}
\put(285,155){\line(1,0){5}}
\put(295,155){\line(1,0){5}}
\put(305,155){\line(1,0){5}}
\put(315,155){\line(1,0){5}}
\put(325,155){\line(1,0){5}}
\put(335,155){\line(1,0){5}}
\put(343,155){\line(1,0){2}}
\put(280,160){$(\!x\!+\!a_i\!+\!2m_1\!)$}


\put(45,5){\line(0,1){235}}
\put(120,5){\line(0,1){235}}
\put(195,5){\line(0,1){235}}
\put(270,5){\line(0,1){235}}
\put(345,5){\line(0,1){235}}
\put(420,5){\line(0,1){235}}
\put(43,251) {$\NSzero$}
\put(118,251) {$\NSone$}
\put(193,251){$\NStwo$}
\put(268,251){$\NSthree$}
\put(343,251) {$\NSfour$}
\put(418,251) {$\NSfive$}


\put(266,135){$\otimes$}

\put(227,135){$(\!x\!+\!m_1\!)$}


\put(231,165){\framebox(5,5){$\cdot$}}
\put(306,105){\framebox(5,5){$\cdot$}}

\put(215,172){$(\!x\!+\!M_j\!)$}
\put(272,96){$(\!x+\!2m_1\!-\!M_j\!)$}

\put(10,30){\vector(1,0){20}}
\put(10,30){\vector(0,1){20}}
\put(05,50){$x$}
\put(30,25){$t$}

\end{picture}
\end{center}
\label{figurethree}
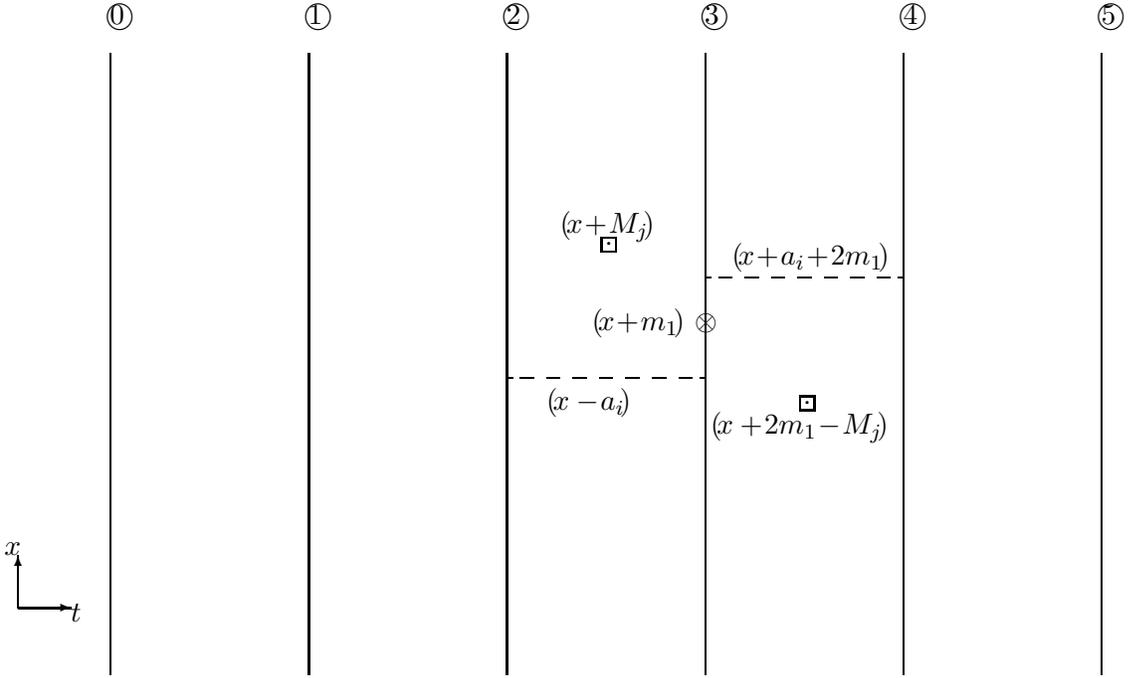
{\footnotesize{\bf Figure 3}: The $m_2\to \infty$ limit of Figure 2. 
In this limit, only the 5-branes $\NStwo$, $\NSthree$, and  $\NSfour$ 
remain connected by 4-branes. 
The other 4-branes and O$6^{-}$ planes have ``slid off'' to $x \sim \infty$.}

\eject

\end{document}